\documentclass[journal,twoside,web]{ieeecolor}
\usepackage{changes}
\usepackage{tmi}
\usepackage{url}
\usepackage{cite}
\usepackage{amsmath,amssymb,amsfonts}
\usepackage{mathrsfs}
\usepackage{algorithmic}
\usepackage{graphicx}
\usepackage{textcomp}
\usepackage{subfigure}
\usepackage{booktabs}
\def\BibTeX{{\rm B\kern-.05em{\sc i\kern-.025em b}\kern-.08em
	T\kern-.1667em\lower.7ex\hbox{E}\kern-.125emX}}
\begin{document}
\title{Quasi-supervised Learning for Super-resolution PET}
\author{Guangtong Yang, Chen Li, Yudong Yao, \IEEEmembership{Fellow, IEEE}, Ge Wang, \IEEEmembership{Fellow, IEEE}, and Yueyang Teng\textit{*}
	\thanks{This work was supported by the Fundamental Research Funds for the Central Universities of China (N180719020).\\
		\indent G. Yang and C. Li are with the College of Medicine and Biological Information Engineering, Northeastern University, Shenyang 110169, China.\\
		\indent Y. Yao is with the Department of Electrical and Computer Engineering, Stevens Institute of Technology, Hoboken, NJ 07030, USA.\\
		\indent G. Wang is with the Department of Biomedical Engineering, Rensselaer Polytechnic Institute, Troy, NY 12180, USA.\\
		\indent Y. Teng is with the College of Medicine and Biological Information Engineering, Northeastern University, Shenyang 110169, China,
		and with the Key Laboratory of Intelligent Computing in Medical Image, Ministry of Education, Shenyang 110169, China.
		(email: tengyy@bmie.neu.edu.cn).}
}
	
\maketitle
	
\begin{abstract}
Low resolution of positron emission tomography (PET) limits its diagnostic performance. Deep learning has been successfully applied to achieve super-resolution PET. However, commonly used supervised learning methods in this context require many pairs of low- and high-resolution (LR and HR) PET images. Although unsupervised learning utilizes unpaired images, the results are not as good as that obtained with supervised deep learning. In this paper, we propose a quasi-supervised learning method, which is a new type of weakly-supervised learning methods, to recover HR PET images from LR counterparts by leveraging similarity between unpaired LR and HR image patches. Specifically, LR image patches are taken from a patient as inputs, while the most similar HR patches from other patients are found as labels. The similarity between the matched HR and LR patches serves as a prior for network construction. Our proposed method can be implemented by designing a new network or modifying an existing network. As an example in this study, we have modified the cycle-consistent generative adversarial network (CycleGAN) for super-resolution PET. Our numerical and experimental results qualitatively and quantitatively show the merits of our method relative to the state-of-the-art methods. The code is publicly available at \url{https://github.com/PigYang-ops/CycleGAN-QSDL}.
\end{abstract}
	
	
\section{Introduction}
\label{sec:introduction}
\IEEEPARstart{P}{ositron} emission tomography (PET) has been instrumental in clinical applications and research studies especially for oncology \cite{b1}, neurology \cite{b2,b3} and cardiology \cite{b4}. However, due to its hardware limitations, PET resolution is typically lower than that of other major imaging techniques \cite{b5,b6} such as computed tomography (CT) and magnetic resonance imaging (MRI). This has reduced its impact on diagnosis and therapy. Therefore, it is of great significance to achieve PET super-resolution computationally.

To date, many super-resolution techniques have emerged. Early super-resolution methods simply utilized interpolation theory \cite{b7,b8} such as nearest-neighbor interpolation, bilinear interpolation and cubic interpolation. Although these methods are fast, the reconstructed images suffer from textural blurring. To address this challenge, image super-resolution as a classical problem in computer vision has attracted a lasting attention, since it is practically valuable and yet inherently ill-posed. Hence, a prior must be used to regularize a super-resolution result. Compressed sensing theory based on sparsity \cite{b9,b10,bbb} has been widely used for super-resolution models. For example, Ren \emph{et al.} \cite{b11} proposed a joint sparse coding (JSC) method to regularize PET images with high-resolution (HR) PET images and anatomical CT or MRI images acquired from the same patient to train a joint dictionary and recover rich details.
	
In recent years, deep learning \cite{b12,b13,ba,bc} has demonstrated extraordinary capabilities in image processing tasks including super-resolution, denoising, etc. Song \emph{et al.} \cite{b16} applied a convolutional neural network (CNN) to improve PET resolution, aided by anatomical information.
Dong \emph{et al.} \cite{b14} proposed a super-resolution CNN (SRCNN) network that learns an end-to-end mapping between low-resolution (LR) and HR images directly, exhibiting faster and better performance than traditional methods. Shi \emph{et al.} \cite{b15} directly extracted feature maps in the original LR space and achieved better performance than SRCNN by introducing an efficient subpixel convolution layer in the network. Lai \emph{et al.} \cite{b17} designed a pyramid network based on cascaded CNNs to achieve more refined mappings. To produce more realistic and detailed images, Ledig \emph{et al.} \cite{b18} introduced a supervised generative adversarial network (GAN) with a new perceptual loss. A common limitation of these approaches is that they are all in the supervised learning mode; i.e., they require paired LR and HR samples which are at an extremely high cost.
	
Unsupervised deep learning uses unpaired data, without requiring paired data. Bulat \emph{et al.} \cite{b19} proposed a two-stage GAN-centered approach. On one hand, a high-to-low GAN learns how to degrade HR images. On the other hand, a low-to-high GAN uses these paired LR/HR images to achieve super-resolution. With the development of cycle-consistent generative adversarial networks (CycleGANs) \cite{b20}, many methods have emerged to improve image resolution through this cyclic refinement. For instance, Yuan \emph{et al.} \cite{b21} and Kim \emph{et al.} \cite{b22} proposed a cycle-in-cycle GAN. The inner cycle maps LR images to pseudo-clean LR images, then performs super-resolution on the intermediate images using a pretrained depth model, and finally fine-tunes the two modules in an end-to-end manner in the outer cycle. Under the condition that no external examples are available for training, Shocher \emph{et al.} \cite{b23} proposed a small image-specific CNN that exploits a cross-scale internal recurrence of image-specific information and solves the super-resolution problem for this specific image by training the model using examples extracted from the test image itself. Furthermore, Song \emph{et al.} \cite{b24} proposed a self-supervised super-resolution method based on dual GANs to learn the map from LR to HR PET images. This method does not require paired data.

\begin{figure}[htb]
	\centering
	\includegraphics[width=\linewidth]{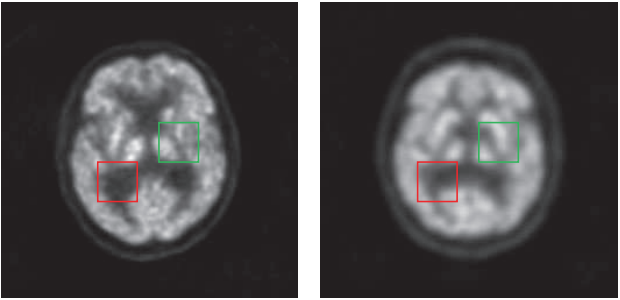}
	\caption{Head PET images of two patients scanned on different scanners. Similarities between the two images are marked with the boxes in the same color (red or green).}
	\label{fig1}
\end{figure}

These unsupervised learning methods use either GAN to distinguish synthetic images from true images or compromise HR images by some prior information. Neither of them harvest all information hidden in a large amount of unpaired data from different patients on various PET scanners in hospitals, resulting in a huge waste of resources. Evidently, even if two images come from different patients, they still have similarities useful to improve a deep learning model. The idea is based on the fact that in the field of medical imaging patient tissues are similar in terms of function and structure, and the PET images should also be similar, even imaged under different conditions on diverse image equipment. Fig. \ref{fig1} shows two PET head images from two patients imaged on different scanners, and it can be seen that there is local similarity in these images. This similarity is important as a source of information to improve deep learning methods.
	
This paper proposes a new weakly-supervised learning mode, which we call quasi-supervised learning, to employ the local similarity among unpaired LR and HR images for neural network construction. 
The involved network can be built based on a new architecture or an existing deep learning model. In fact, supervised and unsupervised network models can be considered as special cases of our quasi-supervised approach. In this study, CycleGAN is taken as an example to be adapted for demonstration of quasi-supervised learning. The experimental results show its flexibility and effectiveness. 

The rest of this paper is organized as follows. In the next section, we describe our methodology. In the third section, we present our experimental design and results. In the last section, we discuss relevant issues and conclude the paper.
	
\section{Methodology}

The proposed quasi-supervised learning method consists of the following steps:
\begin{enumerate}
	\item Acquire a large number of unpaired LR and HR images of many patients.
	\item Divide them into patches containing meaningful local information.
	\item Find best-matched patch pairs from LR and HR patches according to their similarity.
	\item Train a neural network to learn guided by the matched patch pairs and corresponding matching degrees.
\end{enumerate}
Local information processing and neural network construction are the two main elements of our approach for super-resolution PET. In the following, we give key details.
	
\subsection{Local information matching}
The representation and matching of local information are critically important. The former needs to be in a form that can be accessible to machine, and the latter mimics human perception of similarity with a proper metric. Fortunately, patches extracted from a whole image can often represent local features well, and many metrics measure similarity between two patches.
	
\begin{figure*}[htb]
	\centering
	\includegraphics[width=\linewidth]{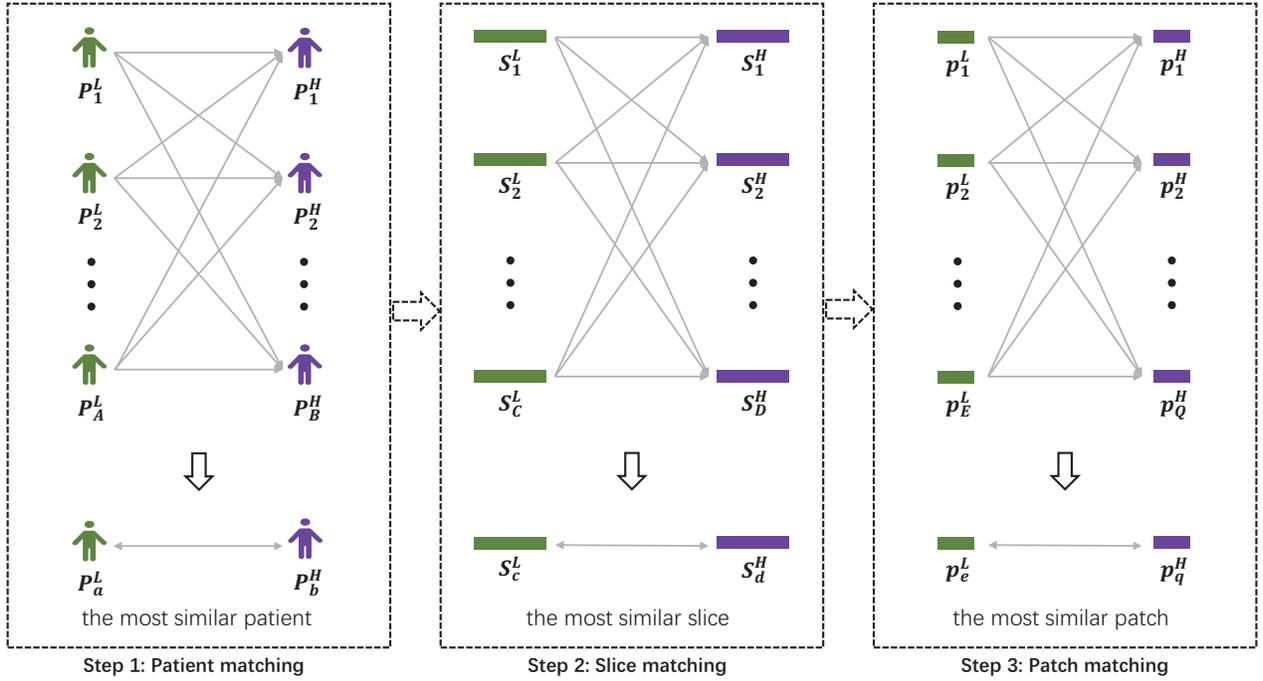}
	\caption{Flowchart for matching LR and HR image patches, where $P_{i}^{L}$/$P_{i}^{H}$, $S_{i}^{L}$/$S_{i}^{H}$ and $p_{i}^{L}$/$p_{i}^{H}$ represent the $i$-th patient, slice and patch in the LR and HR image sets respectively.}
	\label{fig2}
\end{figure*}

For given unpaired LR and HR PET images, our goal is to find the most similar patch pairs in LR and HR respectively and form a large set of such paired patches. Given the large number of patches, direct matching will undoubtedly impose a heavy computational burden. To reduce the computing overhead, we perform the patch matching task at the three levels: patient matching, slice matching, and patch matching. The workflow is shown in Fig. \ref{fig2}, which entails an efficient multi-scale workflow:
\begin{enumerate}
	\item For a patent in an LR image set, find a most similar patient in an HR image set by calculating similarity between the images in the LR and HR image sets.
	\item  For a given slice in the LR image volume, find best-matching slice in the corresponding HR image volume.
	\item For a given patch in the LR slice, find best-matching patch in the corresponding HR slice.
	\item For the corresponding LR and HR patch pairs, compute and store their similarity information.
\end{enumerate}

During the matching process, we can use normalized mutual information (NMI), Pearson correlation coefficient (PCC), radial basis function (RBF), and other functions to measure similarity between two image patches.
	
The calculation of NMI between images $x$ and $y$ can be formulated as
\begin{equation}
	NMI\left (x,y \right ) = \frac{2I\left (x,y \right ) }{H\left ( x \right )+  H\left ( y \right )  }
\end{equation}
where
\begin{eqnarray}
	I\left ( x,y \right ) &=& \sum_{i} \sum_{j}p\left ( x_i,y_j \right )  \log_{}{\frac{p\left ( x_i,y_j \right ) }{p\left ( x_i \right ) p\left ( y_j \right ) } }\\
	H\left (x \right ) &=& - \sum_{i} p\left ( x_{i}  \right ) \log_{}{p\left ( x_{i}  \right ) }
\end{eqnarray}
$x_i$ and $y_j$ are pixel values in images $x$ and $y$, $p\left ( x_i \right )$ and $p\left ( y_i \right )$ denote their distributions respectively, and $p\left ( x_i,y_j \right )$ denotes the joint distribution.
	
PCC is defined as follows: 
\begin{equation}
	PCC \left ( x,y \right ) = \frac{<x,y> }{\sigma _{x} \sigma _{y}}
\end{equation}
where $<x,y>$ is the inner product of $x$ and $y$, and $\sigma _{x}$ and $\sigma _{y}$ are the standard deviations of the pixel values in $x$ and $y$ respectively.
	
RBF is defined as
\begin{equation}
	RBF\left ( x,y \right ) = e^{-\tfrac{\left \| x- y \right \| _{2}^{2} }{2  \gamma ^{2} } }
\end{equation}
where $\gamma>0$ is a given hyperparameter.
	
While our goal is to find the most similar patch pairs, the additional two steps, patient and slice matching, improve the matching efficiency at a cost of potentially reduced accuracy. Therefore, when a dataset is small or computational power is sufficient, these two steps can be removed entirely or partially for the best matching results. The accuracy of data matching certainly depends on the sample size; that is, the larger dataset, the better matching results. The utilization of data in supervised learning involves only two cases, paired and unpaired. In contrast, quasi-supervised learning treats the matching degree as a probability distribution; i.e., the higher the similarity, the higher the probability of matching, which is more aligned with human fuzzy reasoning. 

\begin{figure*}[htb]
	\centering
	\includegraphics[width=\linewidth]{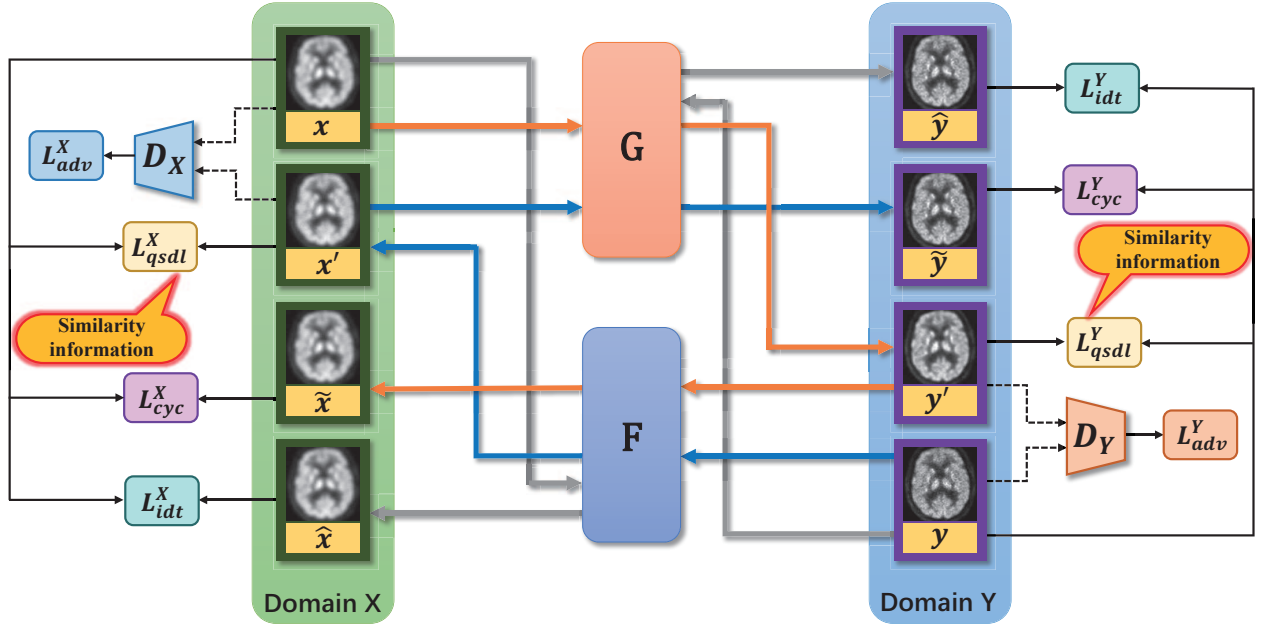}
	\caption{CycleGAN-QL framework for PET super-resolution imaging. Our model contains two generators $G$ and $F$ and two discriminators $D_{Y}$ and $D_{X}$ corresponding to the two generators. Matched LR/HR data and similarity between them are used to train a neural network, which is different from the workflow of the original CycleGAN.}
	\label{fig3}
\end{figure*}
	
\subsection{Neural network construction}

After obtaining a large number of patch pairs and the corresponding similarity measures, the next step is to construct a neural network. An important advantage of the proposed quasi-supervised learning approach is that it can be implemented by modifying existing networks. We take CycleGAN as an example to showcase the proposed method, which is called CycleGAN-QL.

Given the LR and HR image domains $X$ and $Y$, CycleGAN consists of two mappings $G: X\to Y$ and $F: Y\to X$. It also introduces two adversarial discriminators $D_{Y}$ and $D_{X}$, which determine whether the output of a generator is real or not. For example, given an LR image $x\in X$, $G$ learns to generate an HR image ${y}'$ that closely resembles a true HR image $y$ and thus deceives $D_{Y}$. Analogously, $D_{X}$ tries to distinguish the fake output ${x}'$ of $F$ from a true $x$. To regularize the mapping between source and target domains, the network is empowered with four types of loss functions: adversarial loss $\mathcal{L}_{ADV}$, cycle consistency loss $\mathcal{L}_{CYC}$, identity loss $\mathcal{L}_{IDT}$, and quasi-supervised loss $\mathcal{L}_{QL}$.
	
\subsubsection{Adversarial loss}
{The adversarial loss promotes generated images to obey the empirical distribution of the source and target domains. For the mapping $G$ and its discriminator $D_{Y}$, our objective is expressed as
\begin{eqnarray}
	\mathcal{L}_1 \left ( G,D_{Y},X,Y  \right ) = \mathbb{E}_{y\sim p_{data} \left ( y \right ) } \left [ \log_{}{D_{Y} \left ( y \right ) }  \right ]&&\nonumber\\	
	+ \mathbb{E}_{x\sim p_{data} \left ( x \right ) }\left [ \log_{}{\left ( 1- D_{Y} \left ( G\left ( x \right )  \right )  \right ) } \right ]&&
\end{eqnarray}
where $x\sim p_{data} \left ( x \right )$ and $y\sim p_{data} \left ( y \right )$ denote the two data distributions respectively. $G$ aims to minimize the objective, while $D_{Y}$ tries to maximize it. Similarly, the mapping $F$ and its corresponding discriminator $D_{X}$ are formulated as
\begin{eqnarray}
	\mathcal{L}_2 \left ( F,D_{X},Y,X  \right ) = \mathbb{E}_{x\sim p_{data} \left ( x \right ) } \left [ \log_{}{D_{X} \left ( x \right ) }  \right ]&&\nonumber\\	
	+ \mathbb{E}_{y\sim p_{data} \left ( y \right ) }\left [ \log_{}{\left ( 1- D_{X} \left ( F\left ( y \right )  \right )  \right ) } \right ]&&
\end{eqnarray}
Therefore, the total adversarial loss is
\begin{equation}
	\mathcal{L}_{ADV} ( G,F ) =	\mathcal{L}_1  ( G,D_{Y},X,Y ) + \mathcal{L}_2 ( F,D_{X},Y,X)
\end{equation}
}
	
\subsubsection{Cycle consistency loss}
{To further enhance the consistency between the mapping functions, we introduce a cycle consistency loss. This loss can be expressed as
\begin{eqnarray}
	\mathcal{L}_{CYC} \left ( G,F  \right )&=&\mathbb{E}_{x\sim p_{data}\left ( x \right ) }\left [ \left \| F\left ( G\left ( x \right )  \right )  - x\right \| _{1}  \right ]\nonumber\\
&&	+ \mathbb{E}_{y\sim p_{data} \left ( y \right ) }\left [ \left \| G\left ( F\left ( y \right )  \right )  - y\right \| _{1}  \right ]
\end{eqnarray}
where $\left \| \cdot  \right \| _{1}$ denotes the $L_{1}$-norm. Since the cycle consistency loss encourages $F\left ( G\left ( x \right )  \right ) \approx x$ and $G\left ( F\left ( y \right )  \right ) \approx y$, they are referred to as the forward and backward cycle consistencies respectively.
}
	
\subsubsection{Identity loss}
Note that if an HR image is inputted into $G$, the output should be itself, and the same can be said when an LR image is inputted into $F$. Therefore, it is necessary to use the identity loss to guide the model learning. The identity loss can be expressed as
\begin{eqnarray}
	\mathcal{L}_{IDT} \left ( G,F  \right )&=&\mathbb{E}_{x\sim p_{data}\left ( x \right ) }\left [ \left \| F\left ( x \right ) - x \right \| _{1}  \right ]\nonumber\\
&&	+\mathbb{E}_{y\sim p_{data} \left ( y \right ) }\left [ \left \| G\left ( y \right ) - y \right \| _{1}  \right ]
\end{eqnarray}

\subsubsection{Quasi-supervised loss}
The local similarity of unpaired images can be used as an important prior. Our quasi-supervised loss is defined as
\begin{eqnarray}
	\mathcal{L}_{QL} \left ( G,F  \right )=w ( x,y ) \cdot \left \{ \mathbb{E}_{\left ( x,y \right ) \sim p_{data}\left ( x,y \right ) }\left [ \left \| G\left ( x \right ) - y \right \| _{1}  \right ]
	\right.\nonumber\\
	\left.+ \mathbb{E}_{\left ( x,y \right ) \sim p_{data}\left ( x,y \right ) }\left [ \left \| F\left ( y \right ) - x \right \| _{1}  \right ] \right \}
\end{eqnarray}
where $w (x,y)$ represents the weight that are positively correlated with the similarity of $x$ and $y$. This can be seen as the probability of patch matching, with the range of values in $\left [ 0,1 \right ]$. That is, the higher similarity of data pairs, the larger the value of $w\left ( x,y \right )$. Exceptionally, if the training samples are perfectly matched, then $w\left ( x,y \right )= 1$, and the qusai-supervised loss becomes the supervised loss.
	
\subsubsection{Overall loss}
Our overall objective to optimize the two generators $G$ and $F$ and their corresponding discriminators is as follows:
\begin{eqnarray}
	\mathcal{L}_{CycleGAN-QL}&=&  \mathcal{L}_{ADV} \left ( G,F  \right )\nonumber\\
&&	+\lambda _{1}\mathcal{L}_{CYC} \left ( G,F  \right )\nonumber\\
&&	+\lambda _{2}\mathcal{L}_{IDT} \left ( G,F  \right )\nonumber\\
&&	+ \lambda _{3}\mathcal{L}_{QL} \left ( G,F  \right )
\end{eqnarray}
where $\lambda _{1}$, $\lambda _{2}$ and $\lambda _{3}$ are hyperparameters that specify the share of each loss.
Our proposed CyclcGAN-QL model is shown in Fig. \ref{fig3}. 

\subsubsection{$F$, $G$ and $D$ structures}
The structures of networks $F$, $G$ and $D$ are the same as those used in CycleGAN \cite{b25}. First, the generators $F$ and $G$ have the same structure, which can be decomposed into three parts: feature extraction, transformation and reconstruction. The feature extraction part represents a given image as feature maps for transformation. Then, the transformation part produces the final feature maps for the reconstruction part to generate an output image. Specifically, the feature extraction part consists of six convolutional layers,  each of which is followed by an instance norm module and a leaky ReLU activation module. The transformation part uses five residual block modules. The reconstruction part consists of two deconvolutional layers and four convolutional layers, all of which are followed by instance normalizaiton and leaky ReLU activation. Additionally, there is one convolutional kernel in the last convolutional layer, which is the output of the network. The size of all convolutional kernels in the network is $3\times 3$.
		
In the discriminator, each convolutional layer is followed by an instance norm module and a leaky ReLU activation module except for the last two fully connected layers. The first fully connected layer has 1,024 units followed by leaky ReLU activation. Since we use the least square loss between the estimated  and true images, no sigmoid cross entropy layer is applied. Only a fully connected layer with one unit is used as the output. Similar to the generator, we apply convolutional kernels of size $3\times 3$ in all convolutional layers. The number of convolutional kernels in each layer is 32, 32, 64, 64, 128, 128, 256 and 256 respectively.

\subsection{Extension}
We have described CycleGAN-QL in detail, which suggests a good extensibility of the proposed method. In fact, many supervised deep learning methods mapping an LR image $x$ to a paired HR image $y$ can be formulated as
\begin{equation}
	min \sum_{x\in X,y\in Y} Dist(F(x),y)
\end{equation}
where $Dist$ quantifies the discrepancy between $F(x)$ and $y$. We can always modify it to meet quasi-supervised learning as follows:
\begin{equation}
	min \sum_{x\in X,y\in Y} w ( x,y )Dist(F(x),y)
\end{equation}
	
For unsupervised deep learning, this formulation can be considered an additional loss to the original loss. In that sense, our approach covers both supervised and unsupervised deep learning.

\begin{figure*}[htbp]
	\centering
	\subfigure[]{
		\begin{minipage}[t]{\columnwidth}
			\centering
			\includegraphics[width=\columnwidth]{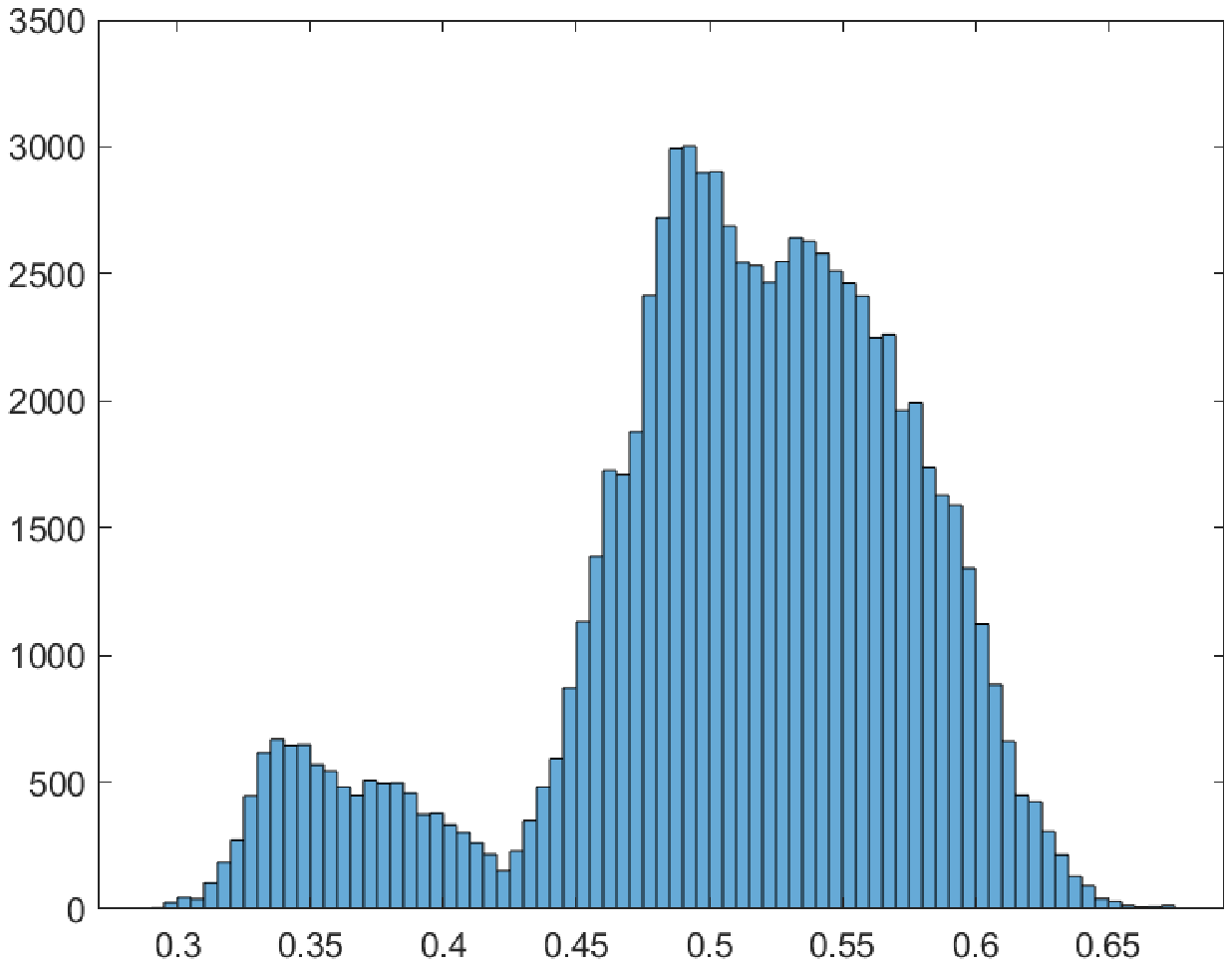}
			\label{fig4}
		\end{minipage}%
	}%
	\subfigure[]{
		\begin{minipage}[t]{\columnwidth}
			\centering
			\includegraphics[width=\columnwidth]{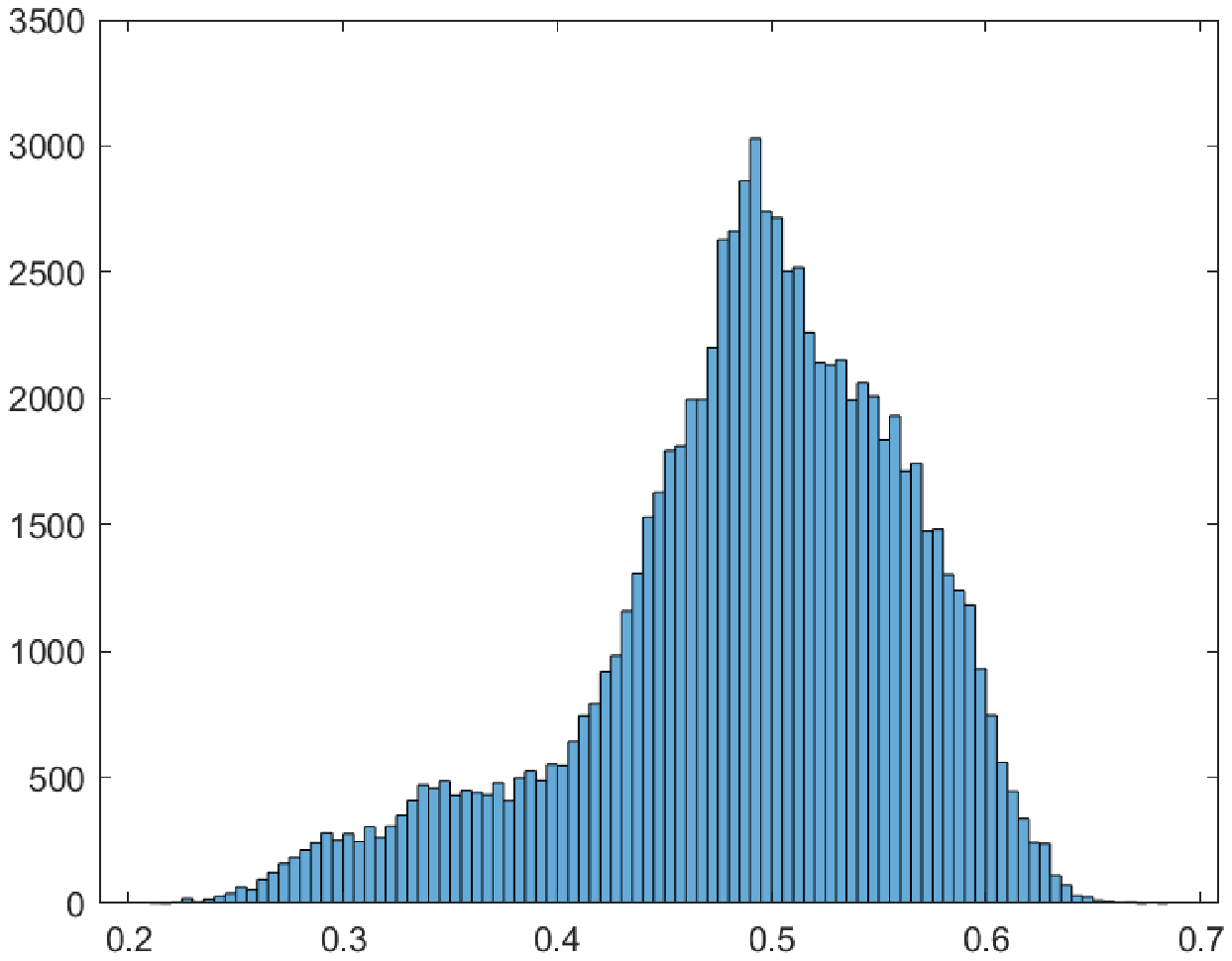}
			\label{fig5}
		\end{minipage}%
	}%
	\centering
	\caption{NMI distributions of patch pairs. The horizontal axis represents NMI, while the vertical axis represents number of patch pairs in size of $128\times 128$: \subref{fig4} the distribution of NMI for truely paired patches, and \subref{fig5} that for best-matched patch pairs.}
	\label{fig100}
\end{figure*}
	
\begin{figure*}[htb]
	\centering
	\subfigure[]{
		\begin{minipage}[t]{\columnwidth}
			\centering
			\includegraphics[width=\columnwidth]{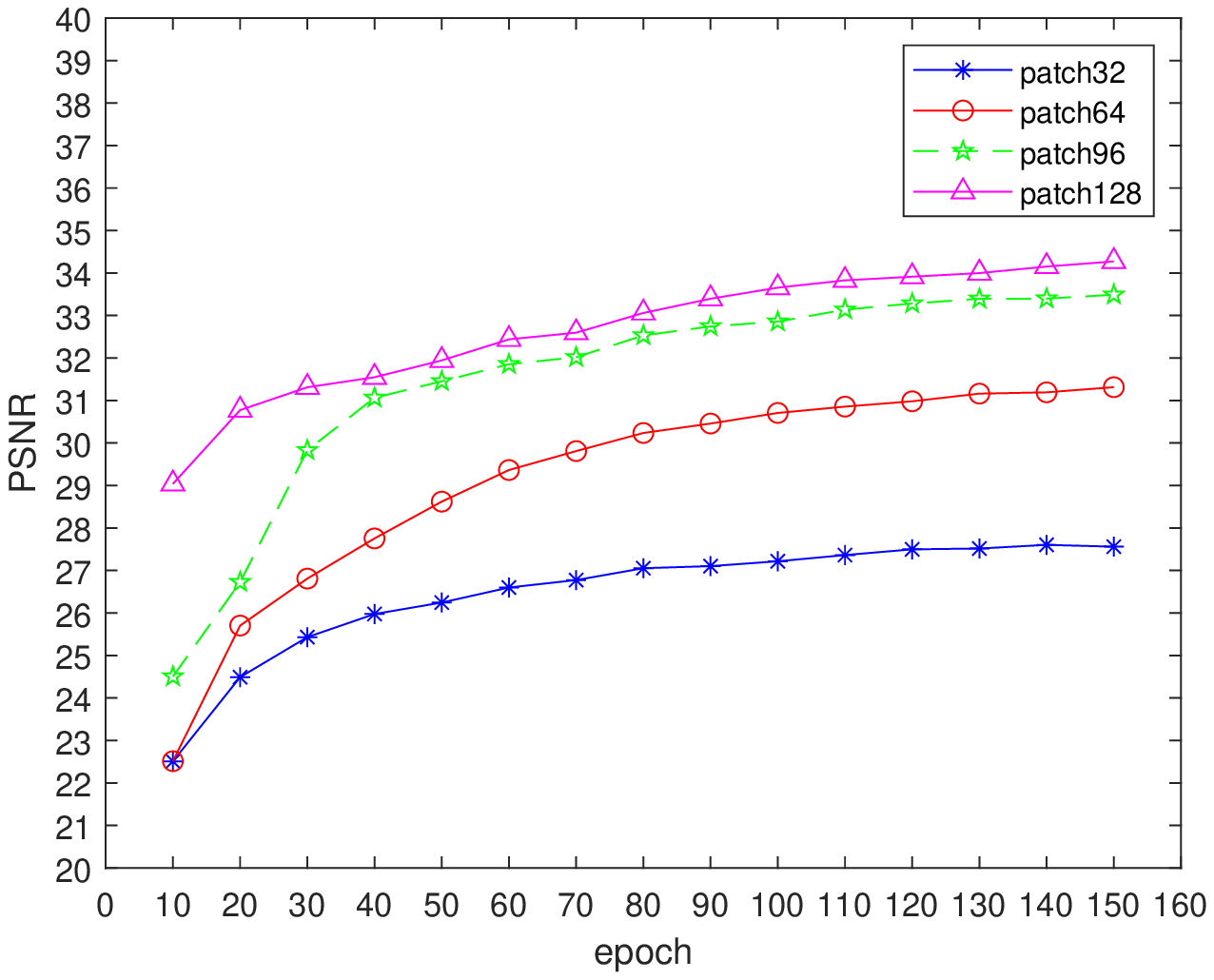}
			\label{fig6}
		\end{minipage}%
	}%
	\subfigure[]{
		\begin{minipage}[t]{\columnwidth}
			\centering
			\includegraphics[width=\columnwidth]{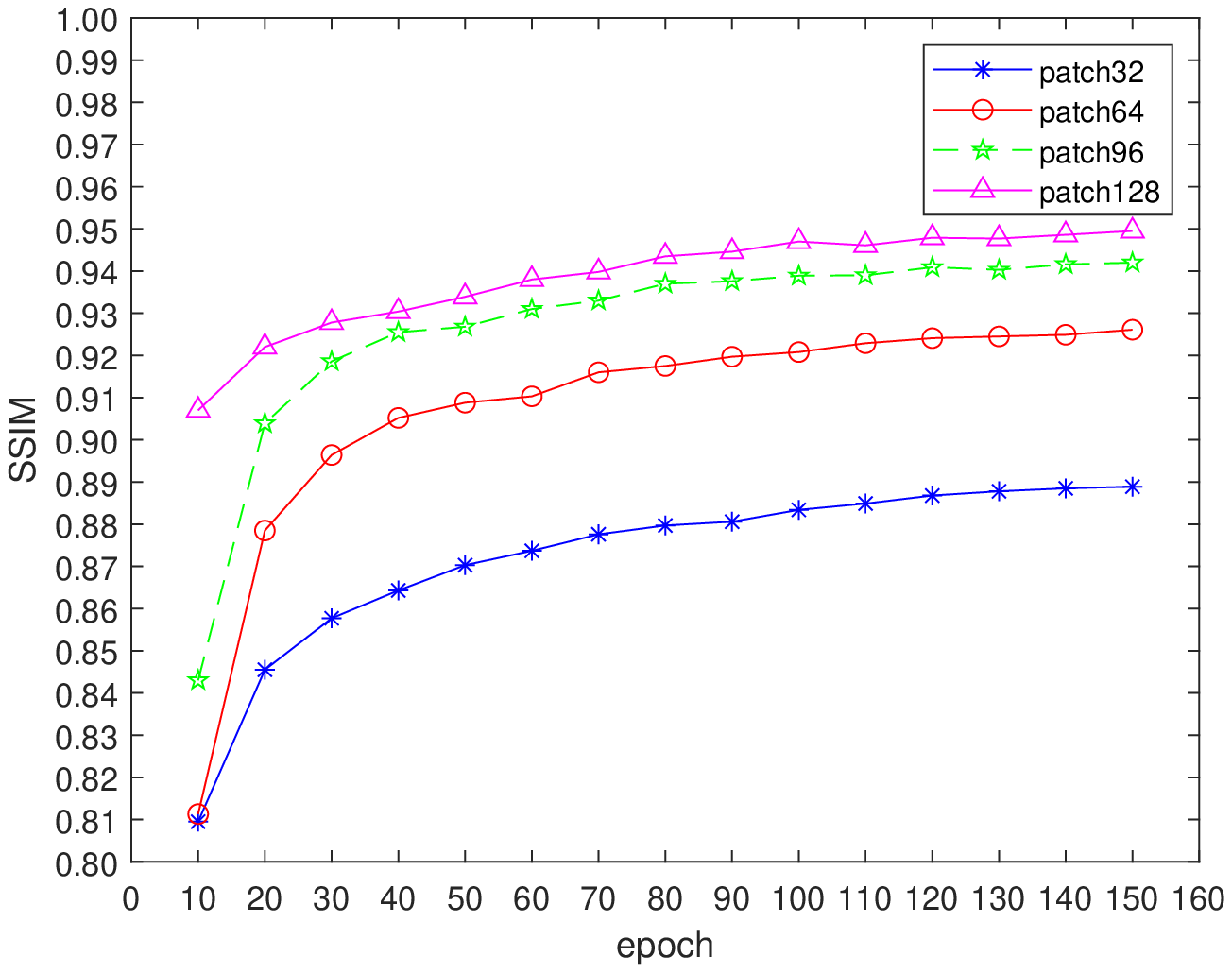}
			\label{fig7}
		\end{minipage}%
	}%
	\centering
	\caption{Effects of training patch sizes on quality metrics: \subref{fig6} PSNR and \subref{fig7} SSIM.}
	\label{fig200}
\end{figure*}
	
\section{Experimental design and results}

To evaluate the performance of quasi-supervised learning, we designed experiments based on the Brain PET Image Analysis and Disease Prediction Challenge dataset \cite{b26}. The dataset consists of 15,000 brain PET images of $128\times 128$ and $168\times 168$ pixels from elderly volunteers. We pre-processed all the PET images in the following procedure. First, we re-sampled all images to the same size of $256\times 256$. Since some images in the dataset had off-center target positions and irregular rotations, which could affect the accuracy of data matching, we performed rotation correction and re-centering for the images. Then, we divided the dataset into two equally sized groups, in which one group was blurred in three steps: an appropriate Gaussian kernel ($\sigma = 3$) blurring, sub-sampling via bicubic interpolation by a factor of 4, and finally up-sampling to the original size. The other group was taken as the HR set, which is the learning target.
	
Since quasi-supervised learning is based on patches, we cropped the images into patches. We extracted overlapping patches from the LR and HR PET images with pre-determined sliding sizes, which allows better preservation of local details. For example, to obtain patches of $128\times 128$ pixels, we set the sliding size to 64, and each image was eventually cropped into 9 patches.
	
In the data-matching phase, we used NMI to quantify the similarity between two images or patches. Since NMI must be in $\left [ 0,1 \right ]$, we directly used it as the weight coefficient; i.e., $w\left ( x,y \right )=NMI(x,y)$.
	
To quantitatively evaluate the performance of the proposed method, we used two widely used image quality metrics: peak signal-to-noise ratio (PSNR) \cite{b27} and structural similarity (SSIM) \cite{b28}. Let $x$ and $y$ denote the true and estimated images respectively. PSNR is defined as follows:
\begin{equation}
	PSNR\left ( y,x \right )= 20\log_{10}{\left ( \frac{\max \left ( y \right )  }{RMSE\left ( y,x \right ) }  \right ) }
\end{equation}
where the root mean square error (RMSE) is defined as
\begin{equation}
	RMSE\left ( y,x \right ) = \sqrt{\frac{1}{n} {\textstyle \sum_{i= 1}^{n}} \left ( y_{i} - x_{i}  \right ) ^{2} }
\end{equation}
	
SSIM is defined as
\begin{equation}
	SSIM\left ( x,y \right ) = \frac{\left ( 2\mu _{x}\mu _{y} +  c_{1}  \right )\left ( \sigma _{xy} +  c_{2}  \right ) }{\left ( \mu _{x}^{2} + \mu _{y}^{2} +  c_{1}  \right ) \left ( \sigma  _{x}^{2} + \sigma _{y}^{2} +  c_{2} \right ) }
	\label{eq7}
\end{equation}
where $\mu _{x}$ and $\mu _{y}$ are the means of $x$ and $y$, $\sigma _{x}^{2}$ and $\sigma _{y}^{2}$ are the variances of $x$ and $y$ respectively, $\sigma _{xy}$ is the covariance between $x$ and $y$, and $c_{1}$ and $c_{2}$ are two parameters stabilizing the division operation.
	
We used the Xavier initializer to initialize the parameters of the convolutional layers \cite{b29}. Considering the computer hardware used, the training batch size was set to 4. As far as the hyperparameters are concerned, we empirically set $\lambda _{1} = \lambda _{2} =1$. We used the Adam optimizer \cite{b30} to optimize the loss function with a learning rate of $0.00001$, $\beta _{1} = 0.5$ and $\beta _{2} = 0.99$. All leaky ReLU activation functions had a slope of $\alpha = 0.2$. All experiments were carried out on a computer with an NVIDIA GeForce RTX 2080 GPU using the TensorFlow library.
	
\subsection{Matched patch pair quality}
Fig. \ref{fig100} shows histograms of truely paired data and our matched data. There is an interesting phenomenon in which NMIs of truely paired data are unexpectedly low, which provides an opportunity to use unpaired data instead of paired data for deep learning. It can be seen that the distribution of the matched data is close to a Gaussian distribution, and most of the matching values are concentrated between 0.45 and 0.55, which may be limited by the dataset scale. When using a larger dataset, the matching values should be higher. In fact, as seen from the experiments below, such degrees of similarity are sufficient to yield a decent performance of the proposed method. Since some patch pairs with very low similarity only provide limited information but greatly increase the computational cost, they need to be removed. In this study, we only kept the patch pairs with similarity greater than the predetermined threshold ($0.4$).
	
\subsection{Hyperparameter selection}
Patches are containers of local information in images, and the patch size may affect the fidelity of local information. Hence, it is necessary to select the patch size properly. We obtained a set of patch pairs in sizes of $32\times 32$, $64\times 64$, $96\times 96$ and $128\times 128$ pixels respectively. For each patch size, we used 20,000 pairs as the training set and 2,500 pairs as the testing set. Fig. \ref{fig200} displays the relationship between patch size and quality metrics with different epoch numbers. PSNR and SSIM was improved when the patch size was increased. Thus, in the subsequent experiments we used $128\times 128$ patches to reflect local features of images. Given our computational resource and favorable image quality already achieved, we have not evaluated the use of image patches larger than $128\times 128$.
	
\begin{figure}[htb]
	\centering
	\includegraphics[width=\linewidth]{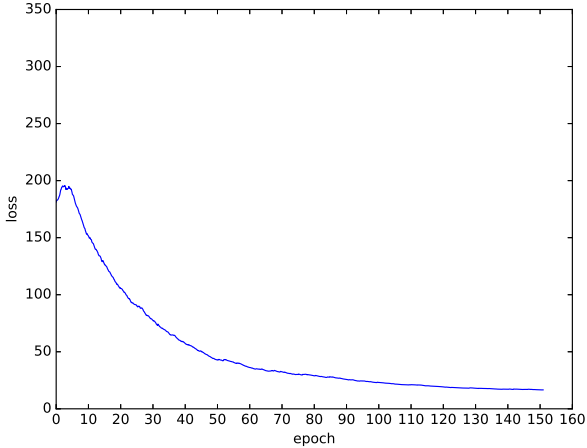}
	\caption{Loss curve on the training set, showing that the loss curve became quite flat about 150 epochs.}
	\label{fig8}
\end{figure}

Fig. \ref{fig8} shows the convergence of the loss curve of the network in the training process. This meets the expectation that the curves converged with iteration. Furthermore, the curve became quite flat about 150 epochs. Based on this observation and to avoid unnecessary training, we set the number of epochs to 150 in the experiments.
	
\begin{figure*}[htb]
	\centering
	\includegraphics[width=\linewidth]{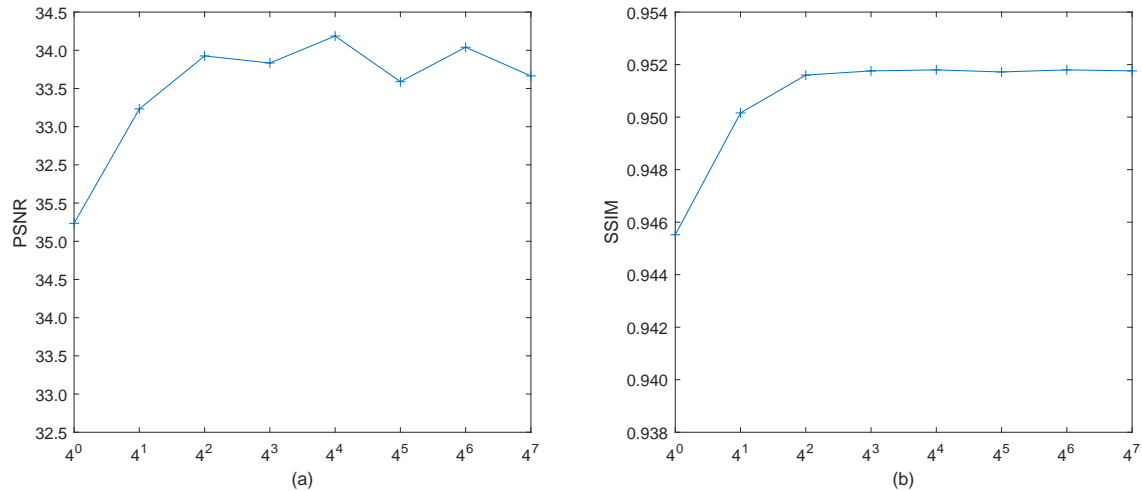}
	\caption{Performance of CycleGAN-QL with respect to the hyperparameter $\lambda _{3}$: (a) PSNR and (b) SSIM.}
	\label{fig9}
\end{figure*}

The quasi-supervised loss is a key to weakly supervised learning. Let us discuss how it is controlled by the weight $\lambda_3$ in the overall loss function.
We will select the optimal weight in the range of $\{4^0,~4^1,...,4^7\}$.
 As shown in Fig. \ref{fig9}, the PSNR curve varies greatly, and the best PSNR was reached when $\lambda _{3} = 4^{4}$. On the other hand, the SSIM curve is very flat after $\lambda_3=4^2$. In fact, most favorable SSIM values were identified when $\lambda_3=4^4$ and $4^6$. Based on a balanced consideration, we set $\lambda _{3} = 4^{4}$ in our experiments and produced satisfactory results.
	
\begin{figure*}[htb]
	\centering
	\includegraphics[width=0.9\linewidth]{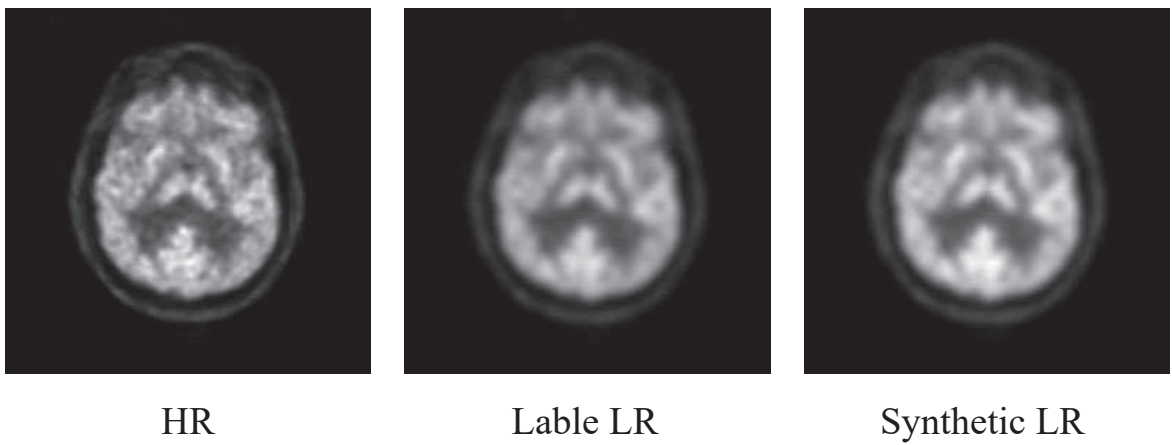}
	\caption{HR, true LR and synthetic LR images.}
	\label{fig10}
\end{figure*}
	
\begin{figure*}[htb]
	\centering
	\includegraphics[width=0.9\linewidth]{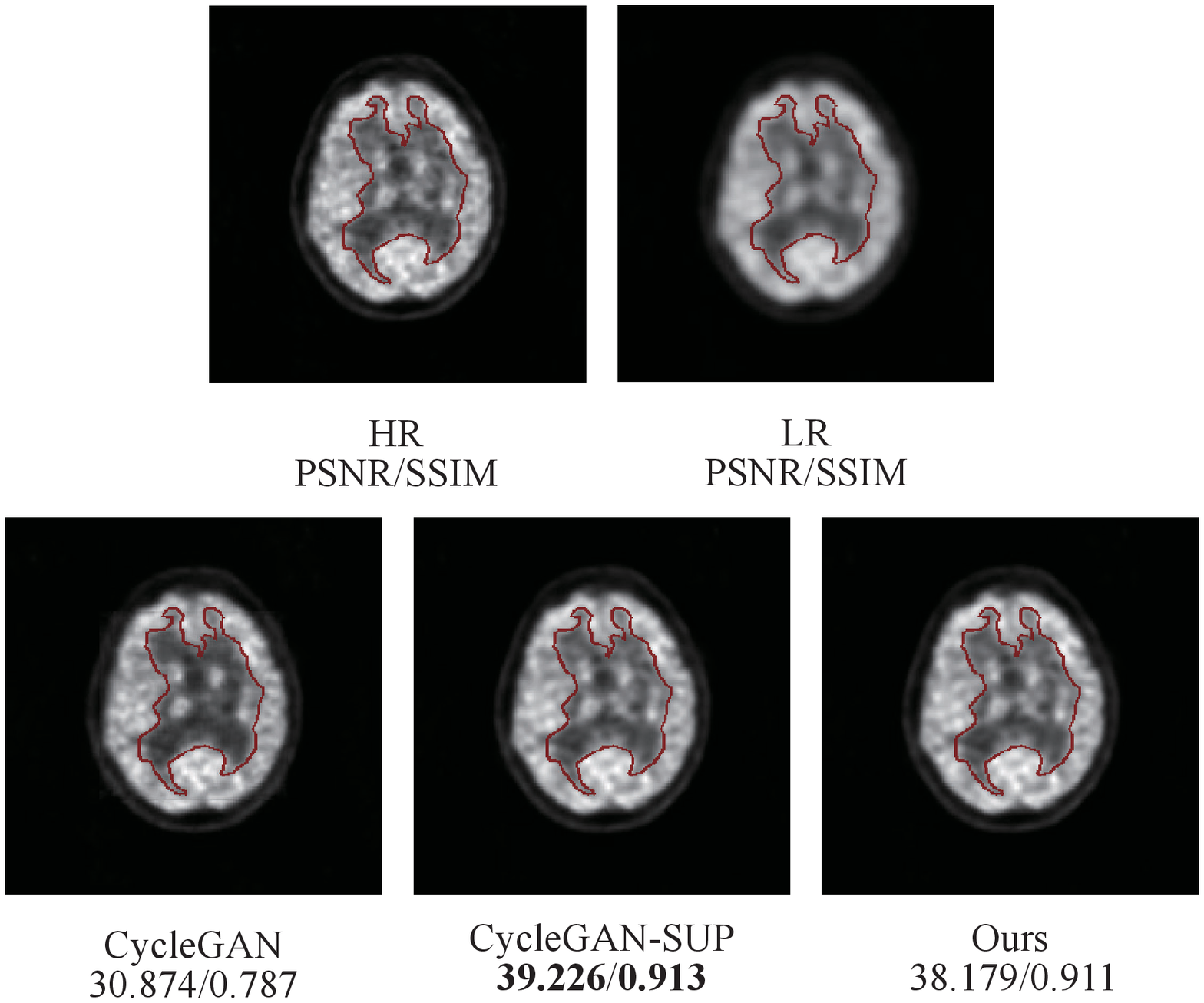}
	\caption{Comparison of CycleGAN-QL variants, where the ventricle are indicated with red boundaries.}
	\label{fig11}
\end{figure*}
	
\subsection{LR image synthesis}
In addition to the super-resolution module, CycleGAN-QL supports resolution degrading, which transforms an HR image to an LR counterpart. Clearly, this process has an implicit effect on resolution improvement. The transformations of HR-to-LR and LR-to-HR are two sides of the CycleGAN mechanism. We show a representative synthetic LR image in Fig. \ref{fig10}. The image looks natural and realistic, and clinical features of the image are kept faithfully, despite image blurring.
	
\subsection{Ablation study}
The proposed CycleGAN-QL can be divided into the following two parts:
\begin{equation}
	CycleGAN-QL=CycleGAN + patch~matching
\end{equation}
To understand the contribution of each part, we evaluated two variants of the proposed method under the same setting: unsupervised and supervised CycleGANs. Each can be considered a special case of CycleGAN-QL:
\begin{enumerate}
	\item Unsupervised CycleGAN, i.e., $\lambda_3=0$ for CycleGAN-QL;
	\item Supervised CycleGAN (named CycleGAN-SUP) with perfectly paired data and unit weight.
\end{enumerate}
	
Fig. \ref{fig11} shows a typical reconstructed slice with the associated PSNR and SSIM displayed. As shown in the figure, both CycleGAN-SUP and CycleGAN-QL deblurred images are very close to the HR image, which are difficult to distinguish visually. The former is  numerically slightly better than the latter. On the other hand, the image obtained with the unsupervised CycleGAN is very different from the HR image, especially within the ventricle as a low-count region where the difference between the HR image and the unsupervised CycleGAN image is far greater than those of the other two methods. The same conclusion can be drawn from the displayed PSNR and SSIM values. 
 
We quantitatively evaluated these methods on the whole testing set in Table \ref{tab1}. The proposed method has a significant advantage over unsupervised CycleGAN  but has a slight performance loss relative to  supervised CycleGAN. This indicates that the local similarity of unpaired data plays a key role in boosting image super-resolution. Moreover, the comparative results with CycleGAN-SUP show that our method can produce image quality similar to the supervised methods. We hypothesize that our method will be further improved, approaching the CycleGAN-SUP performance, if the size of a training dataset is sufficiently large.
\begin{table}
	\caption{Quantitative results of different network variants, where the bold and underlined numbers represent the best and second-best results, respectively.}
	\label{table}
	\setlength{\tabcolsep}{18pt}
	\centering
	\begin{tabular}{lcc}  
		\toprule   
		Method         & PSNR                     & SSIM                      \\  
		\midrule   
		CycleGAN       & 27.144                   &  0.894                    \\
		CycleGAN-SUP   & \bf{36.197}              &  \bf{0.958}               \\
		CycleGAN-QL  & \underline{35.186}       &  \underline{0.955}        \\ 
		\bottomrule  
	\end{tabular}
	\label{tab1}
\end{table}

\subsection{Performance comparison}
We compared CycleGAN-QL with the state-of-the-art methods including total variation (TV) \cite{b24,b31}, GAN \cite{b32,b33}, ADN (artifact disentanglement network) \cite{b34}, CycleGAN and DualGAN \cite{b35}. TV is a classic algorithm for super-resolution and denoising in the image domain. ADN is an unsupervised deep learning-based method for CT artifact reduction. GAN, CycleGAN and DualGAN are successful unsupervised deep-learning methods for image-to-image translation tasks. Except for TV, the methods were trained directly using the original unpaired data, while the proposed method was trained using matched data. 
	
\begin{figure*}[htb]
	\centering
	\includegraphics[width=\linewidth]{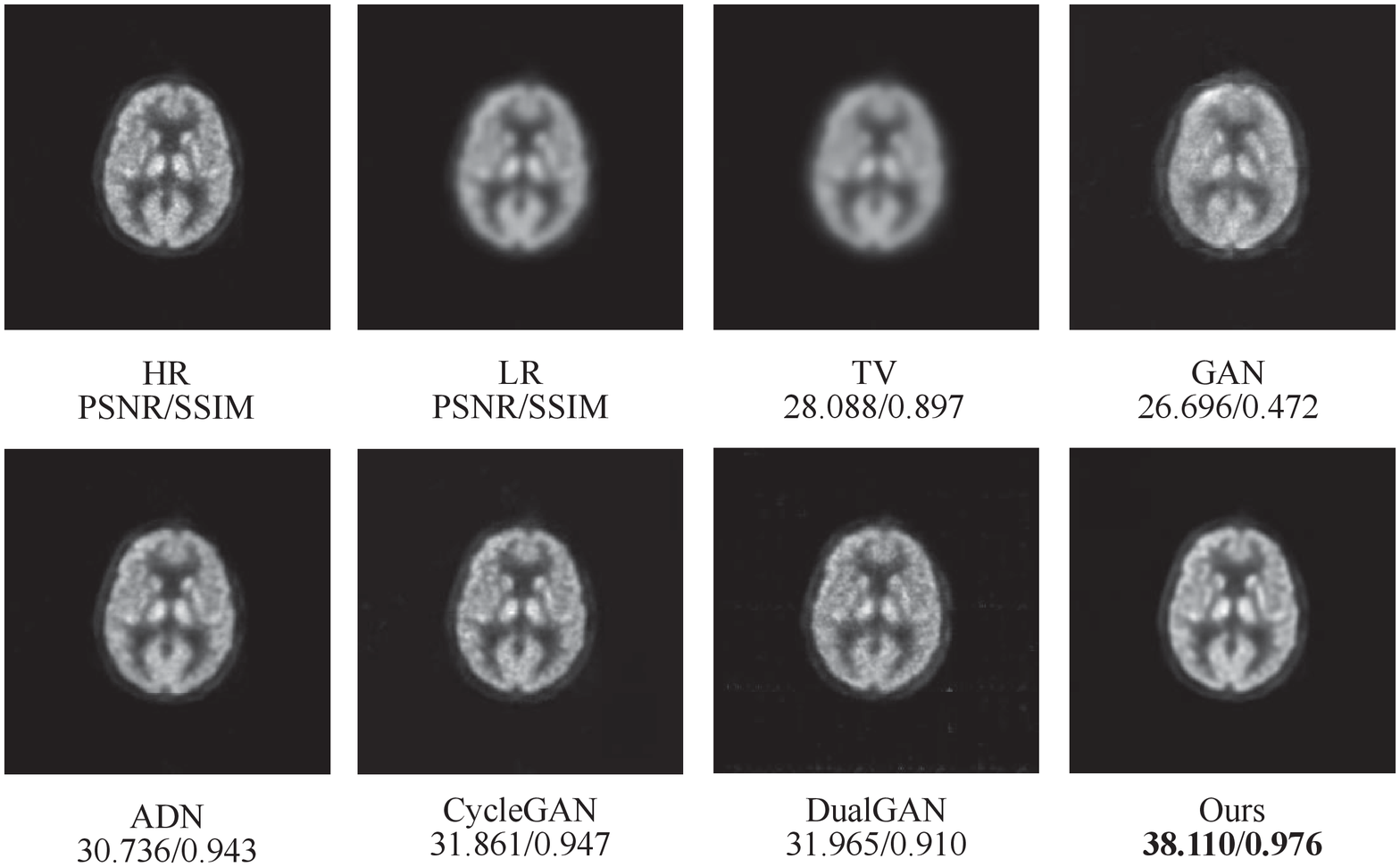}
	\caption{Comparison of the existing and proposed methods for super-resoluton PET.}
	\label{fig14}
\end{figure*}

\begin{figure*}[htb]
	\centering
	\includegraphics[width=\linewidth]{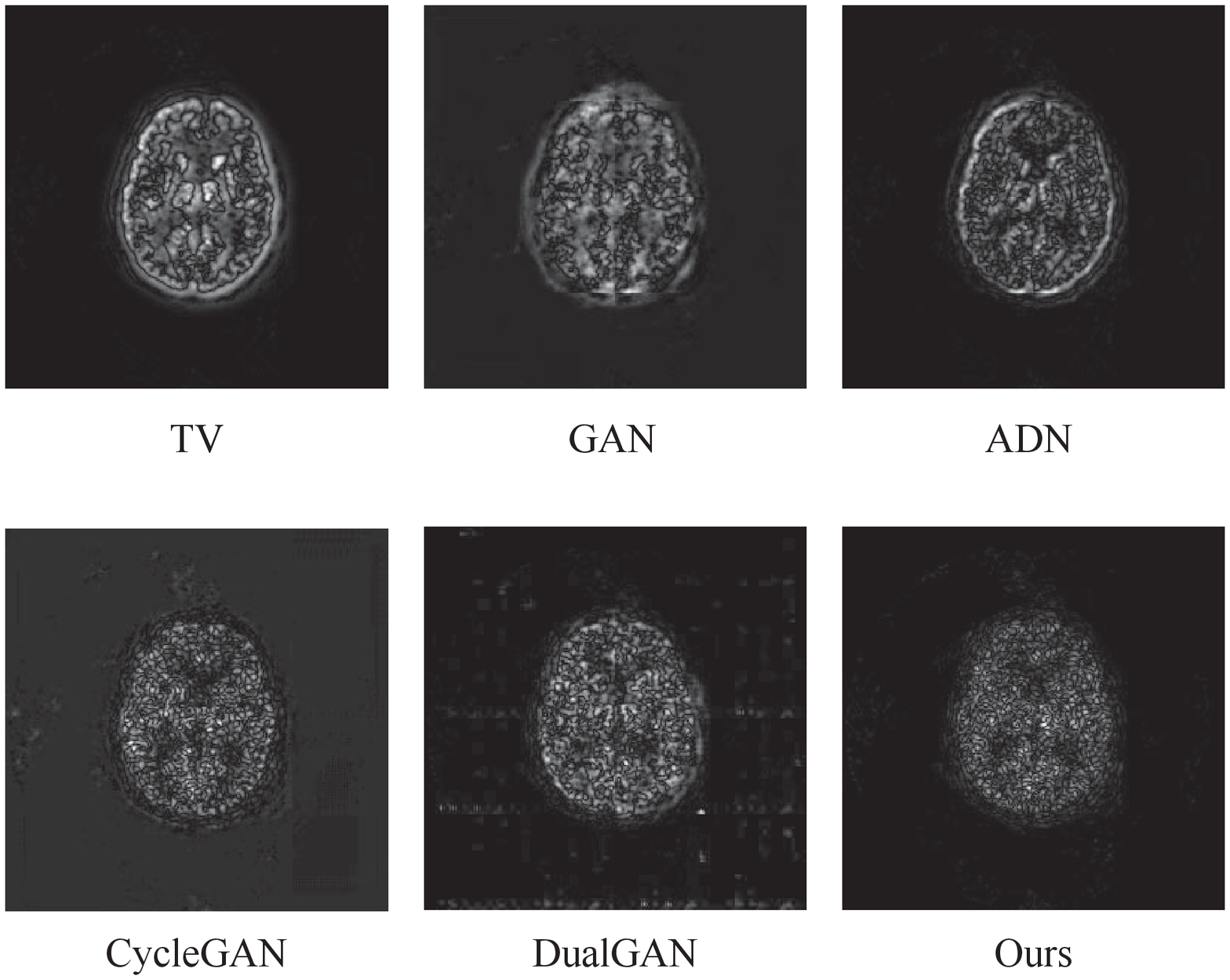}
	\caption{Absolute error relative to the original HR image, which have been scaled to the same gray level.}
	\label{fig15}
\end{figure*}

To qualitatively demonstrate the performance of the proposed method, a representative image is shown in Fig. \ref{fig14}. It can be seen that all the methods improved image resolution of the original image except for TV, which produced blurry results like the LR image. Additionally, DualGAN and CycleGAN, especially GAN, lost feature details in the image. In contrast, our CycleGAN-QL reconstructed the image closest to the true image and recovered most details.
In terms of PSNR and SSIM, the conclusion remains the same.
Additionally, the difference images were obtained by subtracting the generated image from the reference image in Fig. \ref{fig15}, leadning to a conclusion consistent with the above comments.
	
The quantitative results from the various methods on the whole testing set are listed in Table \ref{tab2}. It can be seen that the proposed method produced the images with the highest PSNR and SSIM in general. Although ADN provided the second-best results, significantly below our results, and CycleGAN and DualGAN had very similar performance with ADN. Unexpectedly, GAN presented the worest numerical result, which even is lower than that of TV. In other words, the proposed method can produce a perceptually  pleasing yet quantitatively accurate result, outperforming the other state-of-the-art methods.
	
\begin{table}
	\caption{Quantitative evaluation on the whole testing set using different methods, where the bold and underlined numbers represent the best and second-best results respectively. }
	\label{table}
	\setlength{\tabcolsep}{18pt}
	\centering
	\begin{tabular}{lcc}  
		\toprule   
		Method       & PSNR                     & SSIM                     \\  
		\midrule   
		TV          & 26.896                   &  0.815                   \\  
		GAN         & 12.713                   &  0.561                   \\   
		ADN          & \underline{28.936}       &  \underline{0.897}       \\ 
		CycleGAN     & 27.144                   &  0.894                   \\
		DualGAN     & 28.644                   &  0.854                   \\ 
		Ours         & \bf{35.186}              &  \bf{0.955}              \\ 
		\bottomrule  
	\end{tabular}
	\label{tab2}
\end{table}
	
\section{Discussion and Conclusion}

It is evident that the larger the dataset, the better the diversity of structures and functions a neural network could leverage. Limited by the size of the available dataset and our computational resource, a medium-size dataset was used. In the future, we will consider using more unpaired PET images to improve the super-resolution performance of the network. Also, we need to evaluate the generalizability and stability of this approach, and offer an interpretability of CycleGAN-QL.

Inspired by the fact that images can be mismatched but similar patches in the images can be often found, here we have proposed a quasi-supervised approach to produce PET image super-resolution with unpaired LR and HR PET images. Our method is compatible with either supervised or unsupervised learning, and can be implemented by designing a new network or modifying an existing network, making it applicable in a wider range of applications. This method provides a new way of thinking for computationally upgrading legacy equipment and intelligently developing novel PET scanners. As a matter of fact, the method is not limited to PET image super-resolution and can be extended to other super-resolution and other tasks.

\appendices
	

\begin{IEEEbiography}[{\includegraphics[width=1in, height=1.25in, clip,keepaspectratio]{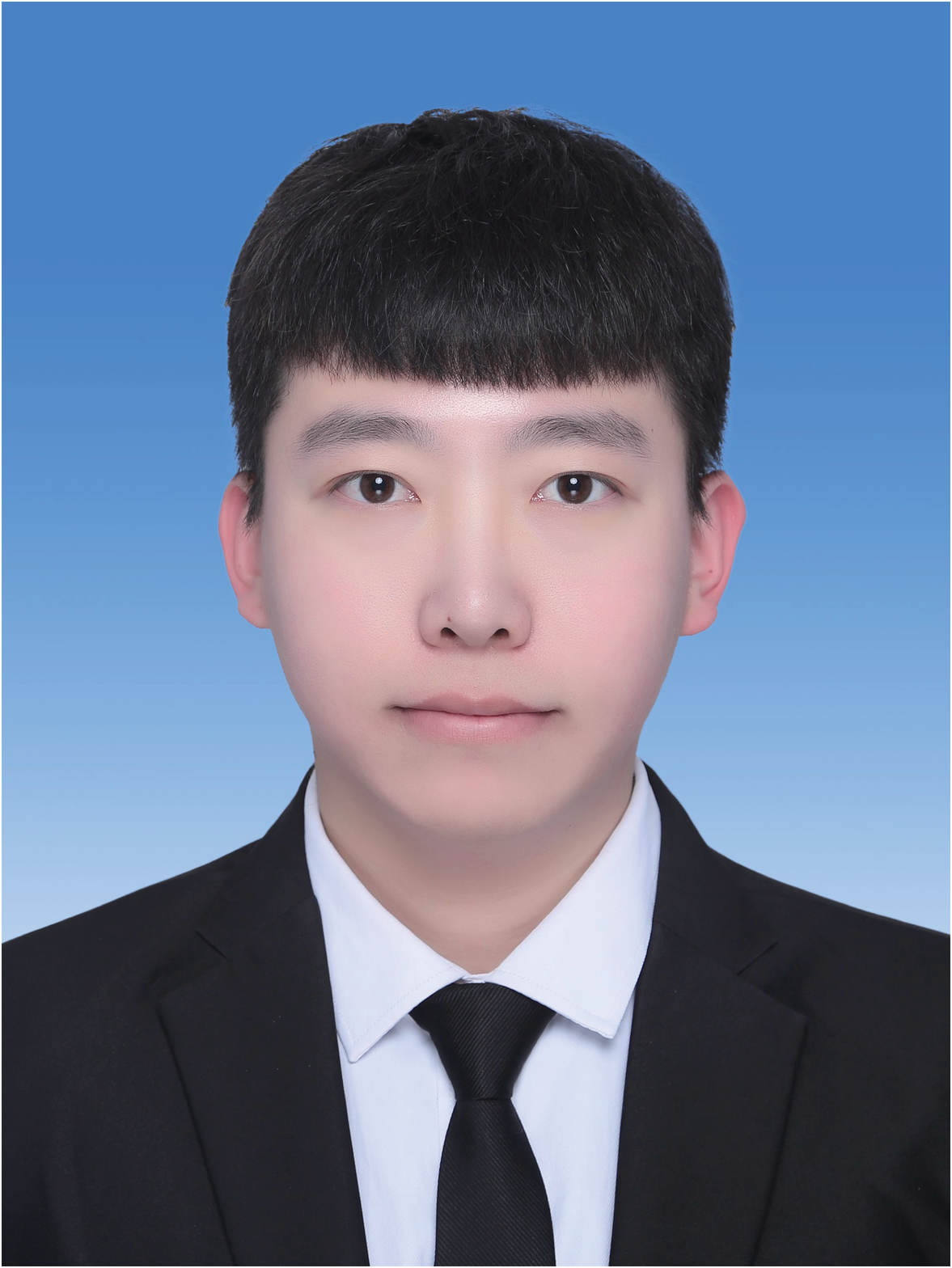}}]{Guangtong Yang} is pursuing his M.S. degree at the School of Medicine and Biological Engineering at Northeastern University and received his B.S. degree from Changchun University of Science and Technology in 2020. His research interests include image processing and deep learning.
\end{IEEEbiography}
\begin{IEEEbiography}[{\includegraphics[width=1in, height=1.25in, clip,keepaspectratio]{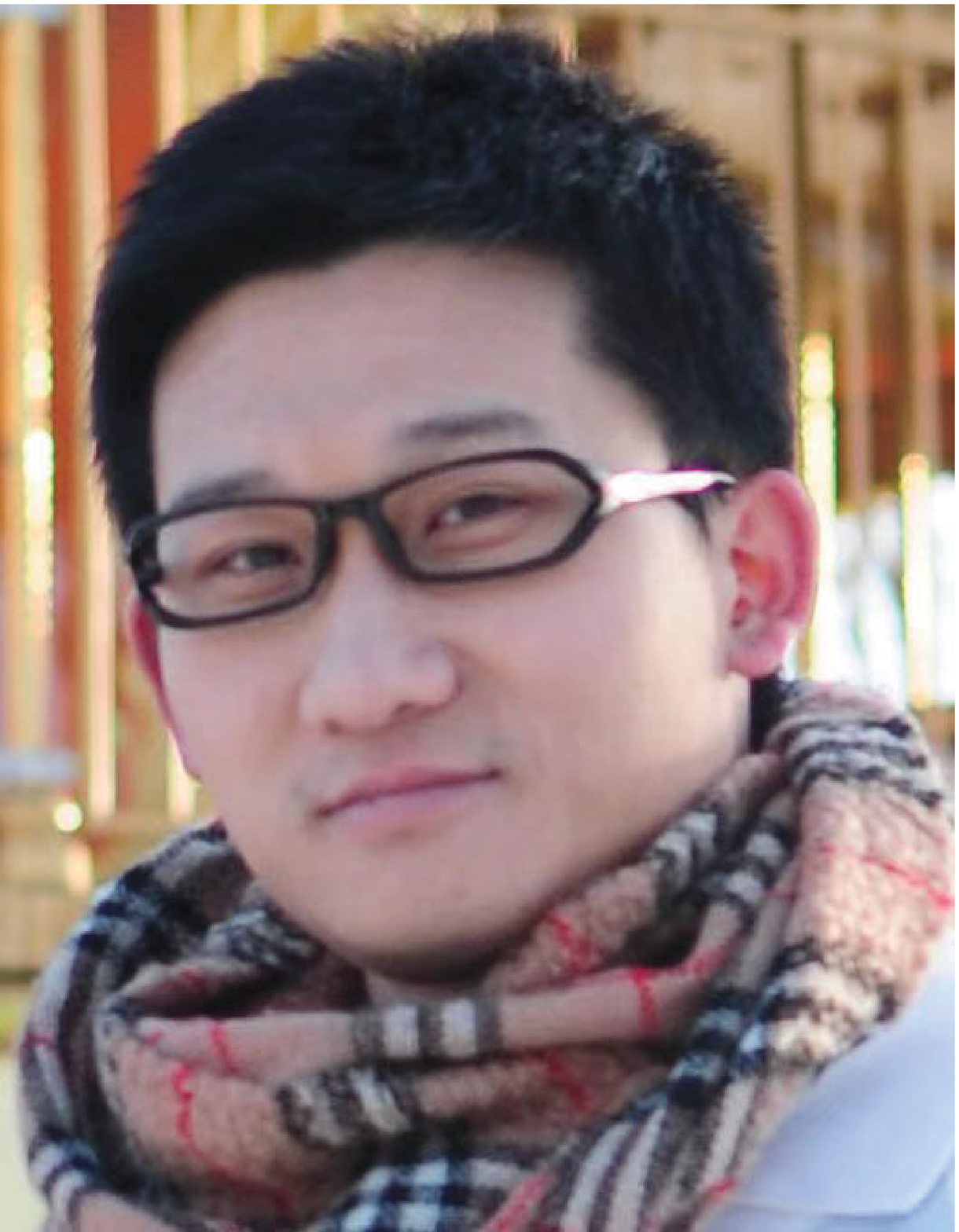}}]{Chen Li} is an associate professor at the College of Medicine and Biological Information Engineering, Northeastern University, Shenyang, China. He is also a guest Ph.D. supervisor at the Institute for Medical Informatics at University of Luebeck, Luebeck, Germany. He obtained his Ph.D. degree in computer science from the University of Siegen in Germany, and completed his postdoc fellow training at the University of Siegen and Johannes Gutenberg University Mainz in Germany. His research interests include pattern recognition, machine learning, machine vision, microscopic image analysis and medical image analysis.
\end{IEEEbiography}
\begin{IEEEbiography}[{\includegraphics[width=1in, height=1.25in, clip,keepaspectratio]{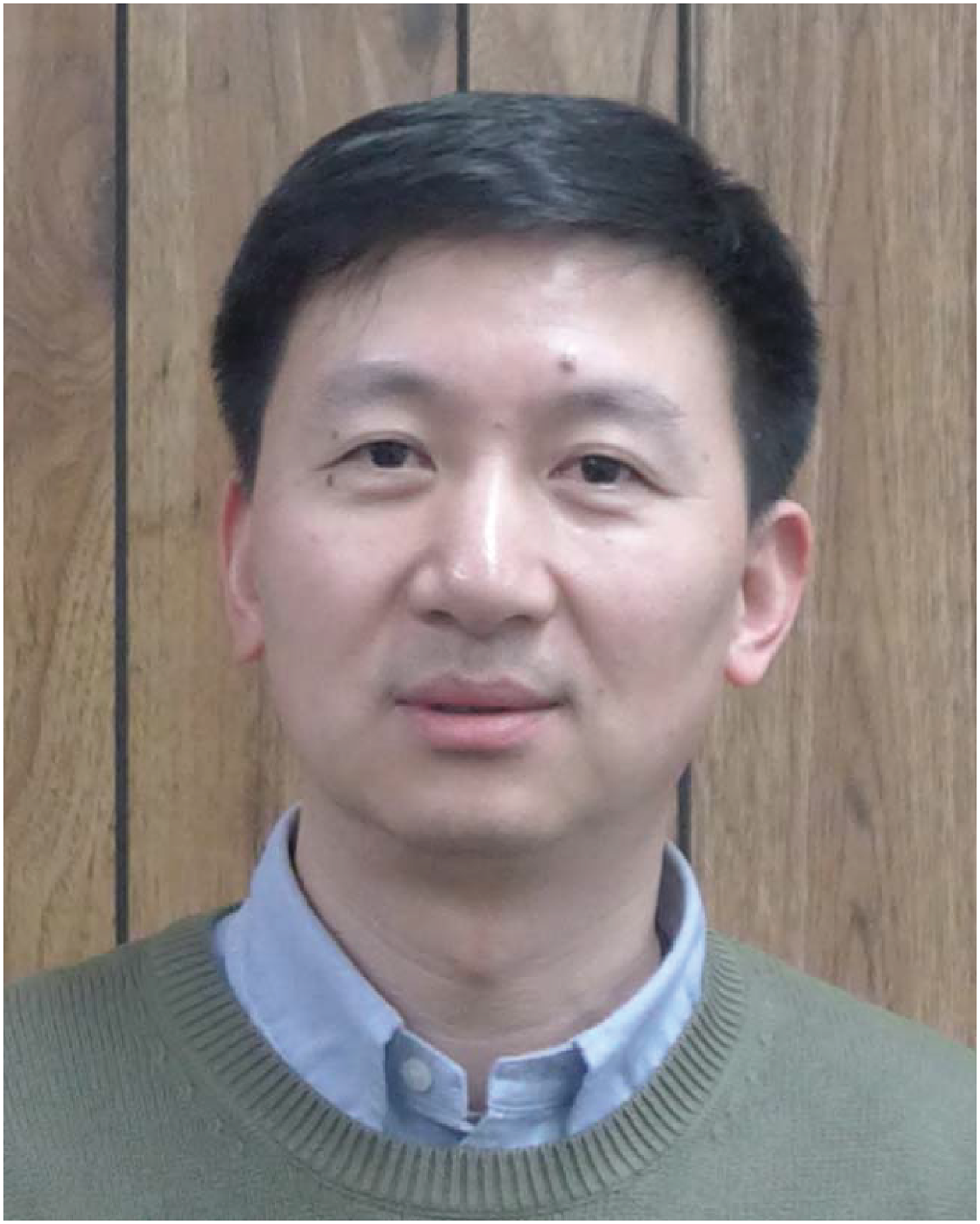}}]{Yudong Yao} (S'88-M'88-SM'94-F'11) received B.Eng. and M.Eng. degrees from Nanjing University of Posts and Telecommunications, Nanjing, in 1982 and 1985, respectively; and a Ph.D. degree from Southeast University, Nanjing, in 1988, all in electrical engineering. He was a visiting student at Carleton University, Ottawa, in 1987 and 1988. Dr. Yao has been with Stevens Institute of Technology, Hoboken, New Jersey, since 2000 and is currently a professor and department director of electrical and computer engineering. He is also a director of Stevens' Wireless Information Systems Engineering Laboratory (WISELAB). From 1989 to 2000, Dr. Yao worked for Carleton University; Spar Aerospace, Ltd., Montreal; and Qualcomm, Inc., San Diego. He has been active in a nonprofit organization, WOCC, Inc., which promotes wireless and optical communications research and technical exchange. He served as WOCC president (2008-2010) and chairman of the board of trustees (2010-2012). His research interests include wireless communications and networking, cognitive radio, machine learning and big data analytics. He holds one Chinese patent and thirteen U.S. patents. Dr. Yao was an associate editor of IEEE Communications Letters (2000-2008) and IEEE Transactions on Vehicular Technology (2001-2006) and an editor for IEEE Transactions on Wireless Communications (2001-2005). He was elected an IEEE Fellow in 2011 for his contributions to wireless communications systems and was an IEEE ComSoc Distinguished Lecturer (2015-2018). In 2015, he was elected a Fellow of the National Academy of Inventors.
\end{IEEEbiography}

\begin{IEEEbiography}[{\includegraphics[width=1in, height=1.25in, clip,keepaspectratio]{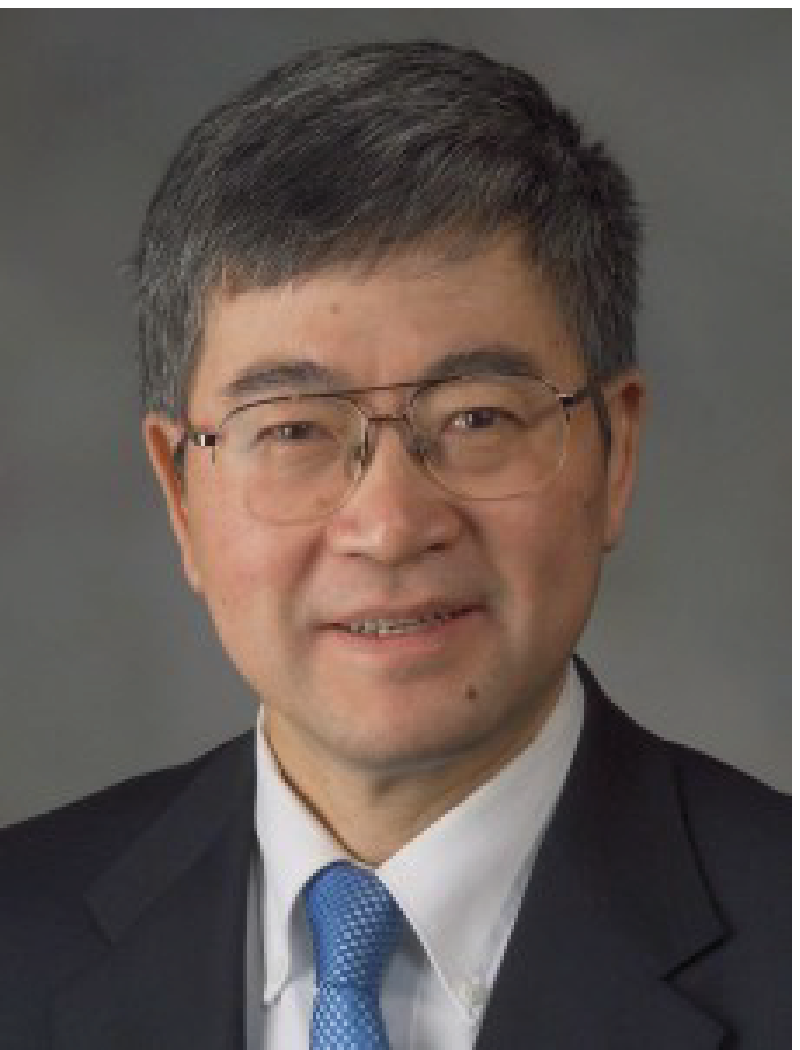}}]{Ge Wang}, PhD in ECE, is Clark \& Crossan Chair Professor and Director of Biomedical Imaging Center, Rensselaer Polytechnic Institute, Troy, New York, USA. He pioneered the spiral/helical cone-beam method in the early 1990s and wrote many follow-up papers in this area. There are ~200 million medical CT scans yearly, a majority of which are performed in the spiral cone-beam mode. He published the first perspective on AI-empowered tomographic imaging in 2016, and a series of papers on diverse deep learning-based imaging topics. He wrote 550+ journal articles in PNAS, Nature, Nature Machine Intelligence, Nature Communications, and other high-quality journals. He holds 100+ issued/pending US or international patents. He has given many seminars, keynotes and plenaries, including NIH AI Imaging Presentations (2018) and SPIE O+P Plenary (2021). He is Fellow of IEEE, SPIE, AAPM, OSA, AIMBE, AAAS, and National Academy of Inventors (NAI). He received various academic awards, with the recent ones including IEEE EMBS Academic Career Achievement Award (2021), IEEE R1 Outstanding Teaching Award (2021), SPIE Aden \& Marjorie Meinel Technology Achievement Award (2022), and Sigma Xi Walston Chubb Award for Innovation (2022).
\end{IEEEbiography}

\begin{IEEEbiography}[{\includegraphics[width=1in, height=1.25in, clip,keepaspectratio]{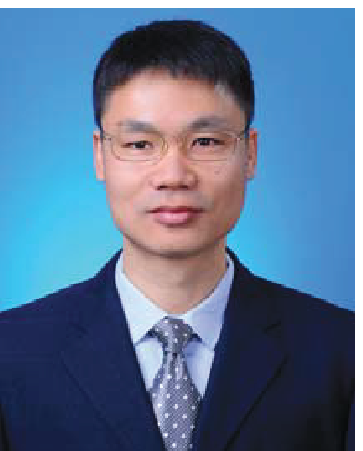}}]{Yueyang Teng} received his bachelor's and master's degrees from the Department of Applied Mathematics, Liaoning Normal University and Dalian University of Technology, China, in 2002 and 2005, respectively. From 2005 to 2013, he was a software engineer at Neusoft Positiron Medical Systems Co., Ltd. In 2013, he received his doctorate degree in computer software and theory from Northeastern University. Since 2013, he has been a lecturer at the Sino-Dutch Biomedical and Information Engineering School, Northeastern University. His research interests include image processing and machine learning.
\end{IEEEbiography}

\end{document}